\documentclass{aa}  

\usepackage{multirow,graphicx}
\usepackage{xcolor}
\usepackage{ulem}
\usepackage{MR_AA}
\usepackage{txfonts}

\newcommand{\tIRFM}{T_\mathrm{IRFM}}
\newcommand{\geff}{g_\mathrm{eff}}
\newcommand{\lapp}{L_\mathrm{app}}
\renewcommand{\Lsun}{L_\odot}
\renewcommand{\Msun}{M_\odot}
\renewcommand{\Rsun}{R_\odot}

\begin{document} 

\title{Fundamental properties of two rapidly rotating stars:\\
Rasalhague and Alkaid}

\titlerunning{Rasalhague and Alkaid}
\authorrunning{Lazzarotto et al.}

\author{Axel Lazzarotto\inst{1}, Alain Hui-Bon-Hoa\inst{1}, Torsten
B\"ohm\inst{1}, Matt Gent\inst{1}  and Michel Rieutord\inst{1}}

   \institute{IRAP, Universit\'e de Toulouse, CNRS, CNES,
14, avenue \'{E}douard Belin, F-31400 Toulouse, France\\
              \email{axel.lazzarotto@irap.omp.eu}
             }

   \date{Received; accepted }

  \abstract
   {The fundamental parameters of rapidly rotating stars are key quantities to understand the impact of rotation on stellar evolution. A few nearby early-type stars offer the possibility of precise measurements of these parameters, which will help us constrain newly available two-dimensional models.
}
   {We propose a method to retrieve the fundamental parameters of a fast rotating star (mass, rotation rate, and age), and also the inclination of its rotation axis on the line of sight, using five spectrophotometric observables along with a set of steady 2D-models.}
   {From two photometric indices, the temperature $\tIRFM$ derived by the infrared ﬂux method, the projected equatorial velocity ($V\sin i$) and the apparent luminosity, along with a grid of 2D steady state models, we select models that are compatible with all observational constraints, and derive the most probable mass, rotation rate, core hydrogen mass fraction relative to that of the envelope, and inclination of the rotation axis on the line of sight of the targetted star.}
   {We apply this method to two stars: Rasalhague ($\alpha$ Oph, A5IV) and Alkaid ($\eta$ UMa, B3V). We confirm and improve the fundamental parameters of Rasalhague previously determined by IR-interferometry and provide a new determination of its rotation axis inclination on the line-of-sight, which we find to be $68.9\pm5.6^{\circ}$. Concerning Alkaid, for which little is known, we infer a mass of $5.071\pm0.023 M_\odot$, a rotation rate of $0.413\pm0.057$ times the critical angular velocity, corresponding to an equatorial rotation period of 14.6 hours. We also find an inclination of the rotation axis of $41.9\pm8.2^{\circ}$ on the line-of-sight. We show that Alkaid is a very young star, presumably between 2 and 8 Myrs off the Zero-Age Main Sequence. As a side result, using high resolution spectra and the Least Square  Deconvolution method, we determined a precise value of the $V\sin i$ of Rasalhague, namely $224.3\pm2.6$~km/s. Similarly, we find  $V\sin i=154.3\pm9.1$~km/s for Alkaid.}
   {Our results show that fundamental parameters of rapidly rotating stars can be determined (at least bracketed) using five observable quantities that can be measured from the spectrum of the star. This is of great interest for stars that are beyond reach of interferometric observations.}


\maketitle

\section{Introduction}

About 50\% of early-type stars exhibit fast rotation (i.e., equatorial
velocity higher than 100~km/s) according to \cite{ZR12}. For such
stars rotational evolution is a key aspect of their global evolution,
driven by nuclear reactions. It is indeed associated with rotational
mixing, which impacts both the surface abundances and the lifetime
of the star by fueling the convective core with hydrogen. Actually,
the interface region between the core and the radiative envelope
is also the place probed by gravity modes, for instance in $\gamma$
Doradus or Slowly Pulsating B stars \cite[e.g.][]{mombarg+22}. Their
oscillation spectrum is therefore much affected by the mixing around it
\cite[e.g.][]{burssens+23}. However, the impacts of processes associated
with rotation are still not well understood. Fast rotation makes them
important and implies the use of two-dimensional models, which can deal
with the oblate shape of the star, the large-scale baroclinic flows, and
anisotropic transport. A step forward was accomplished with the release
of the ESTER code that can compute 2D models including all these effects
\citep{ELR13,RELP16,mombarg+23}. However, as for any model, we now need
precise observational constraints to keep the models on the right tracks,
so as to ensure that its predictions are correct.

Observables derived from stellar spectra have long been key quantities
to infer the fundamental parameters of stars. When rotation is fast,
the spectral energy distribution is more complex to interpret, but also
contains more informations since it is influenced by the so-called gravity
darkening, which makes the polar regions brighter than the equatorial ones
\citep{ELR11}. Provided the use of realistic enough models, it is possible
to tightly constrain the fundamental parameters of a fast rotating
star. This was the challenge taken up by \cite{lazzarotto+23}. Indeed,
taking advantage of gravity darkening, these authors developed a method
that uses photometric quantities to retrieve some fundamental parameters
of a rotating star such as mass and rotation rate together with the
inclination of the rotation axis. They combined ESTER models, PHOENIX
atmospheric structures and synthetic spectra \citep{hauschildt+99}
to produce spectral energy distributions . As a validation test, they
determined the fundamental parameters and inclination of Vega, with a
very good agreement with previous determinations. However, they assumed
an evolutionary status of the star as given by Vega's concordance model
of \cite{monnier_etal12}.

In the present work, we wished to go further and determine the age
of the star. However, we restricted our study to early-type stars
because they own a convective core, whose hydrogen mass fraction can
be used as a proxy of the evolutionary status. In addition, we also
wished to compare our determination of the equatorial radius to that
obtained by interferometric observations, if available. To this end,
we considered two stars: Rasalhague ($\alpha$ Oph) and Alkaid ($\eta$
UMa). Rasalhague\footnote{If not stated otherwise, Rasalhague refers to
the primary of this binary system \cite[e.g.][]{gardner+21}.} is a nearby
star located at 14.8 pc from the Sun \citep{gardner+21} that has been
observed with interferometers \citep{zhao+09,baines+18,baines+25}. Its
mass and rotation rate makes it {\it a priori} similar to Vega but
seen almost equator on, according to \cite{zhao+09}. Hence, expectedly,
gravity darkening makes it of later type than Vega along with a higher
luminosity class, namely A5IVnn \citep{gray+01} versus A0V for Vega
\citep{negueruela+24}. The suffix ``nn'' following the luminosity
class, means ``very broad lines'' \citep[e.g.][]{abt_morrell95},
which is consistent with a rapidly rotating star seen equator-on. It is
therefore an interesting case to test a method based on gravity darkening
effects. The second star, Alkaid, was chosen essentially because its
spectral energy distribution is accurately known from space data and
available through the CALSPEC database \citep{bohlin+22}. As a hot star
\citep[spectral type B3V,][]{negueruela+24}, it also allowed us to test
our method in this range of effective temperatures and, hopefully,
to give new reliable values of its fundamental parameters. We note
that its angular diameter has also been estimated by interferometry
\citep{baines+18,gordon+19,abe+24,baines+25}.

The paper is organised as follows: we first give an overview of
our method in Sect.~\ref{numerics} and then discuss the cases of
Rasalhague and Alkaid in Sect.~\ref{rasalhague} and Sect.~\ref{alkaid},
respectively. Discussion and conclusions follow in Sect.~\ref{conclusion}.

\section{Overview of the method}\label{numerics}

\subsection{Observational constraints}

To constrain the fundamental parameters of the stars, we used five
quantities. The first one was the apparent bolometric luminosity
$L_\mathrm{app}$, derived from the bolometric flux and the distance. We
recall that because of gravity darkening this quantity depends
much on the inclination $i$ of the rotation axis with respect to
the line of sight. From a high resolution spectrum, we derived the
projected equatorial velocity $v\sin i$. Then, the surface effective
temperature $T_\mathrm{IRFM}$ was obtained by the Infra-Red Flux Method
\citep[IRFM,][]{blackwell+80}. Finally, two photometric indices $c'_1$
and $c_2$, defined in \cite{lazzarotto+23} and inspired by Strömgren
$c_1$ photometric index \citep{Stromgren1963} were used. Photometric
index $c'_1$ was defined like \cite{Stromgren1963}'s $c_1$ index
($c_1=(u-v)-(v-b)$) but with an $u$ filter shifted blueward to avoid
unsolved discrepancies between models and observations in the wavelength
range of the actual $u$ filter. This new filter has a central wavelength
of 215~nm. $c_2$ was defined the same way as $c_1$ but with two bandpasses
on the blue side of the Balmer jump and the third one on the red
side. More explicitly, we used the three HST filters HSP\_VIS\_F419N\_B,
ACS\_HRC\_F250W, and HSP\_UV1\_F220W\_A, to define the $c_2$ index as:

\begin{equation}
\label{expression_c2}
c_2=-\frac{5}{2}\left( \log \left( \frac{\int F_\mathrm{F220}}{\int F_\mathrm{F250}} \right)-\log \left( \frac{\int F_\mathrm{F250}}{\int F_\mathrm{F419}} \right)\right)\quad\mathrm{mag}
\end{equation}
These two indices have been designed to reflect the shape of the spectrum around the Balmer jump. As shown in \cite{lazzarotto+23}, they are sensitive to the effects of gravity darkening.

\begin{figure}[t]
\includegraphics[width=\linewidth]{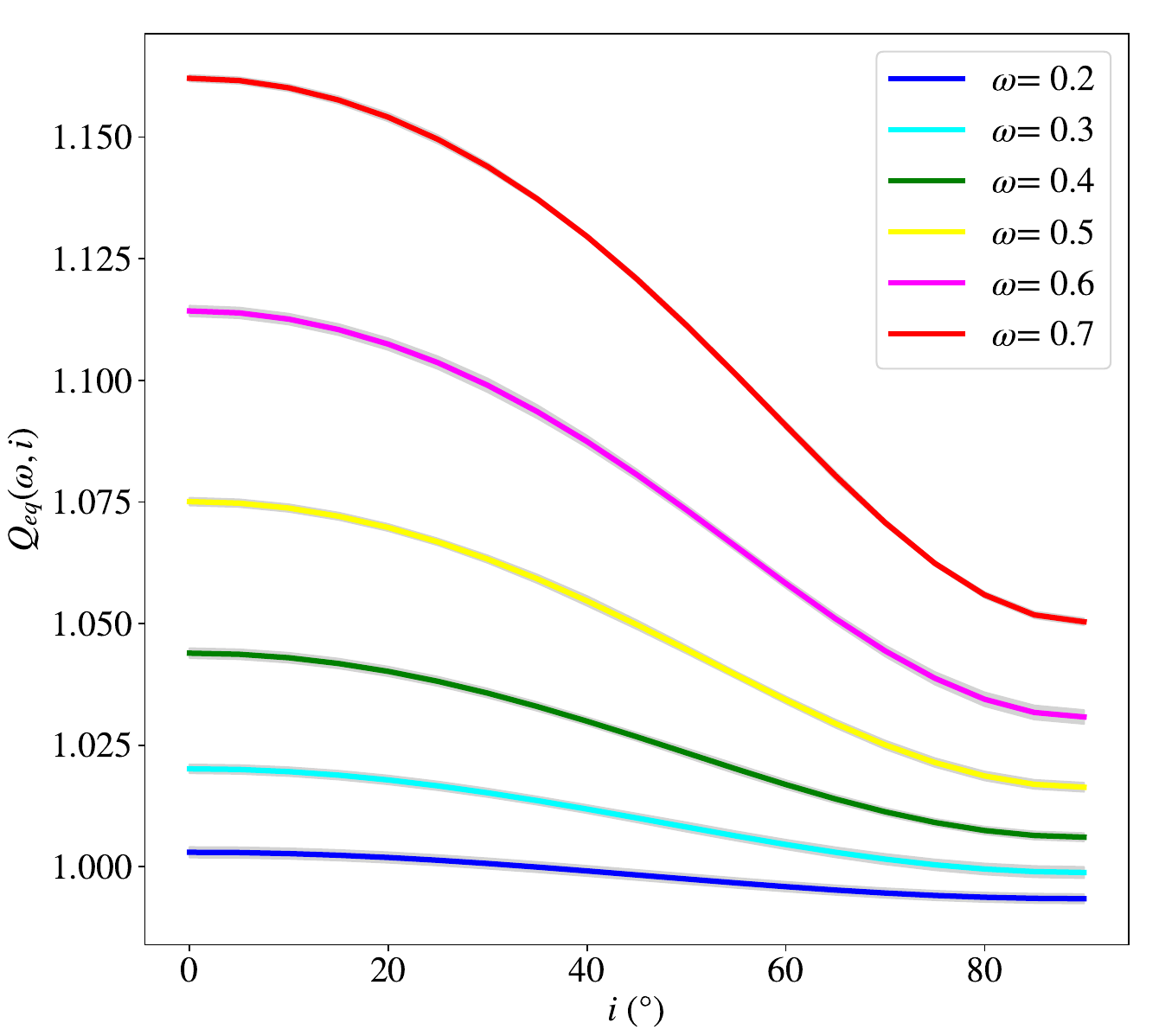}
\caption{Ratio $Q_{eq}(\omega,i)=\tIRFM/T_\mathrm{eff}^{\rm eq}$ as a
function of the rotation axis inclination $i$ on the line of sight, for various rotation rates $\omega$ between 0.2 and 0.7. The thickness of the lines represents the effects of mass and $X_\mathrm{core}/X_\mathrm{env}$ variations in the intervals [2.0,2.4] and [0.25,0.45], respectively.}
\label{Qeq}
\end{figure}

\subsection{Models and spectra synthesis}\label{models}

For the determination of the fundamental parameters of our two stars in focus, we used 2D steady state models computed with the ESTER code\footnote{The code is publicly available at  https://ester-project.github.io/ester.}. This way of doing does not follow the traditional match between a box in the Hertzsprung-Russell diagram and an evolutionary path. In the present state of 2D-models, this is not doable numerically. We therefore considered a grid of steady state models, whose matching with observables provided us with the parameters of the star.

\begin{figure}[t]
\includegraphics[width=1.0\linewidth]{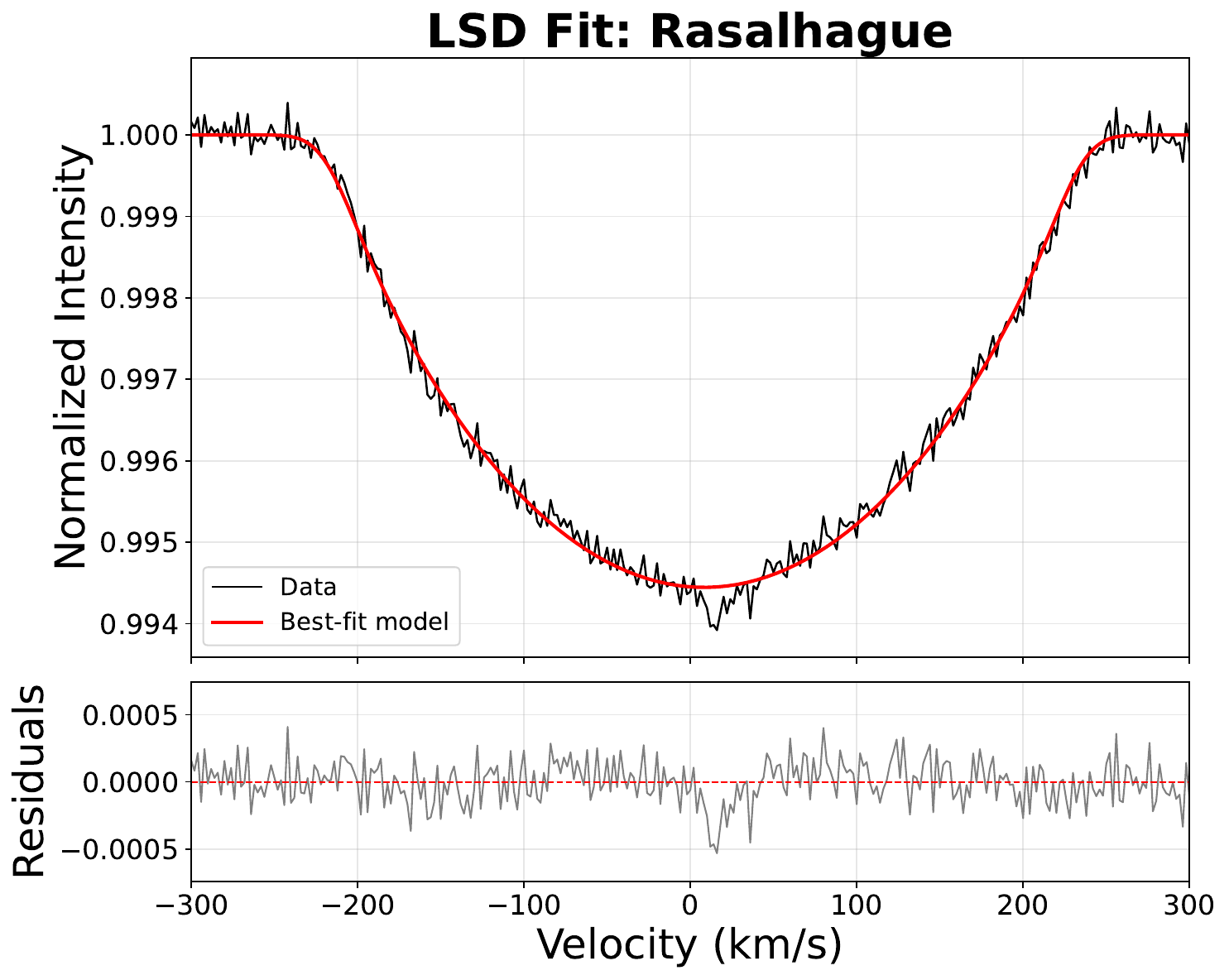}
\caption{Mean line profile of Rasalhague obtained with the Least Square
Deconvolution method \citep{donati+97} with a rotationally broadened profile fit to the data (top panel), and its residuals (bottom panel). The resulting projected equatorial velocity is $224.3\pm2.6$~km/s.}
\label{Rasal_Vsini}
\end{figure}

\begin{figure}[t]
\includegraphics[width=\linewidth]{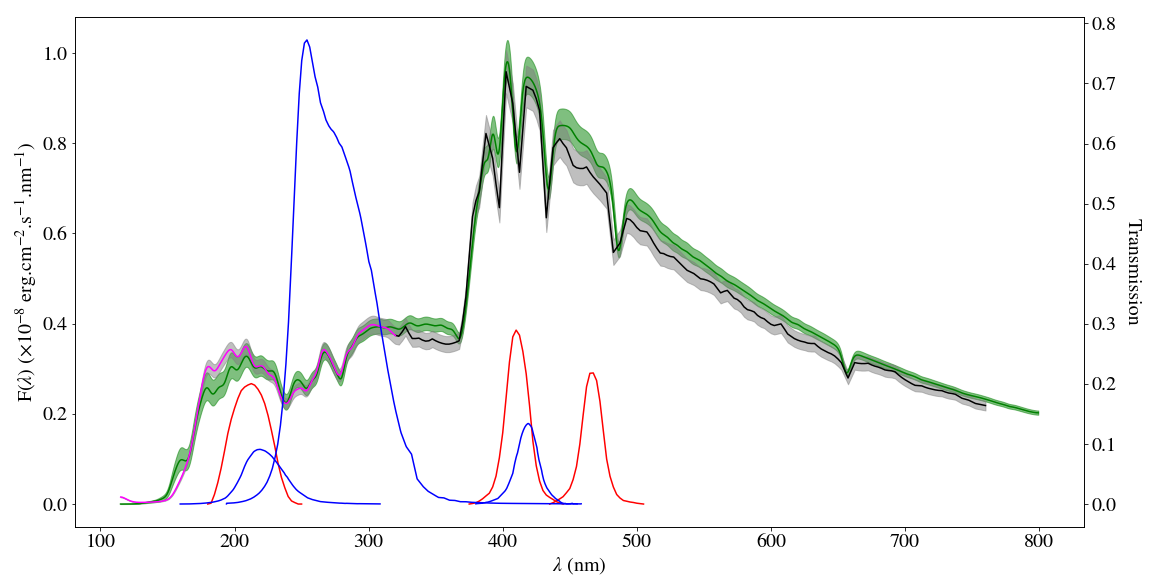}
\caption{Combined spectra of Rasalhague. The spectrophotometric part \#1401 is shown by the black solid line. Its extension towards the UV is in magenta. The 5\%-uncertainty of these data is shown by the greyed zone around the spectrum. The bandpasses of the filters used in the $c'_1$ and $c_2$ indices are shown in red and blue, respectively. The infrared extension is not shown. The green line shows the synthetic spectrum derived from the concordance model given in Tab.~\ref{Final_Rasalhague}. The greened zone around this line shows the effect of uncertainties of the rotation axis inclination.
}
\label{figRasalSpectres}
\end{figure}

We first recall that steady state 2D-ESTER models \citep{ELR13} are defined by five quantities:

\begin{itemize}
    \item the mass $M$,
    \item the hydrogen mass fraction in the core $X_\mathrm{core}$,
    \item the hydrogen mass fraction in the envelope $X_\mathrm{env}$,
    \item the ratio $\omega=\Omega_\mathrm{eq}/\Omega_c$ of the equatorial angular velocity to the critical one (hereafter, rotation rate refers to $\omega$),
    \item the metallicity $Z$.
\end{itemize}
We used $X_c=X_\mathrm{core}/X_\mathrm{env}$ as a proxy of the age. To
simplify the search, we also assumed fixed values for $X_\mathrm{env}$ and
$Z$, which we adjusted to the observed metallicity of the star. Lastly,
we recall that all observed quantities also depend on the orientation $i$
of the rotation axis, which has therefore to be determined.

Once the model was computed, the specific intensity spectra were
calculated with the PHOENIX code \citep{hauschildt+99}. The contribution
of the various surface elements of the visible part of the star were
summed up to yield the visible spectrum, taking into account their
Doppler shift and angle between the normal to their surface and the line
of sight. Hence, gravity darkening and limb darkening effects are included
in the modelling of emergent intensity. Compared to \cite{lazzarotto+23},
we found that the ultraviolet (UV) part of the spectra was poorly sampled
using the 0.1~nm wavelength step, so a finer step of $10^{-3}$~nm was
used shortward of 510~nm. Then, the three photometric quantities $\tIRFM$,
$c'_1$, and $c_2$ were computed as described in \cite{lazzarotto+23}.

The computation of synthetic spectra is intense numerically since it
demands the computation of atmospheric structures on the set of effective
temperatures and effective gravities along the meridian of the star. To
avoid much of these calculations we used interpolating functions over
a grid of pre-computed models adapted to the star we aim to model. For
$\tIRFM$, we actually used a relation between the equatorial effective
temperature $T_\mathrm{eff}^{\rm eq}$ and $\tIRFM$. Indeed, as shown in
Fig.~\ref{Qeq}, the ratio $\tIRFM/T_\mathrm{eff}^{\rm eq}$ essentially
depends on $\omega$ and $i$, and marginally on mass and $X_c$. Thus doing
we could compute several thousands of models with parameters around a
first approximate model of the star under focus.

Among all the grid models, acceptable models were identified using the
following algorithm: for the model under investigation, we derived an
interval of inclinations $i_\mathrm{mod}$ from its equatorial velocity
$v_\mathrm{eq}^\mathrm{mod}$ such that $v_\mathrm{eq}^\mathrm{mod}\sin
i_\mathrm{mod}$ and the observed $v\sin i$ were consistent within the
uncertainties. We then defined the interval of model apparent luminosities
corresponding to this inclination interval. If the so-computed interval
of apparent luminosity had a non-void overlap with the observed one,
the current model was tagged as matching this observable. A similar
procedure was performed with the other observables ($\tIRFM$, $c'_1$, and
$c_2$). If the current model was tagged as matching all the observables,
we selected it as a valid solution and moved on to the next one of the
grid. Since the two stars at hands have also been observed with infrared
(IR) interferometers \citep{zhao+09,baines+18,baines+25}, we also made
an additional test considering the equatorial radius to see how the set
of solution changed when this additional constraint is taken into account.

\begin{table}[t]
\caption{Observational constraints for Rasalhague.}
\label{Rasal_obs}
\begin{center}
\begin{tabular}{cc}
\hline
\hline
$T_\mathrm{IRFM} (\mathrm{K})$&$8040\pm100$  \\
$c_1'$&$1.310\pm0.163$\\
$c_2$& $4.748\pm0.137$\\
$L_\mathrm{app}\ (L_\odot$)&$26.6\pm1.8$\\
$v \sin\ i$ (km.s$^{-1}$)&$224.3\pm2.6$\\
\hline
\hline
\end{tabular}
\end{center}
\end{table}

\begin{table*}[t]
\caption{Parameters of the initial ESTER model of Rasalhague.}
\label{Rasal_param}
\begin{center}
\begin{tabular}{ccccccccccc}
\hline
\hline
$M$ ($M_\odot$) &$R_p$ ($R_\odot$) & $R_e$ ($R_\odot$) & $\epsilon$ & $L$ $(L_\odot)$ & $\teff^p$ (K) & $\teff^e$ (K) & $\log\geff^p$ &
$\log\geff^e$ & $v_\mathrm{eq}$ (km.s$^{-1}$) & $X_{\rm core}/X_{\rm env}$ \\
&&&&&&&&&&\\
2.220 & 2.385 &2.865 & 0.168 & 31.1 & 9183  & 7709 & 4.03 & 3.65 & 242 & 0.37 \\
\hline
\hline
\end{tabular}
\end{center}
\end{table*}

\section{Rasalhague -- $\alpha$ Oph}\label{rasalhague}
\subsection{Observational constraints}

The first constraint we used is the apparent luminosity. From the bolometric flux $f_\mathrm{bol}=3.88\pm0.19\times 10^{-9}$ W/m$^2$ \citep{baines+18} and the distance $d=14.80\pm0.13$~pc \citep{gardner+21}, we derived an apparent luminosity of $\lapp=26.6\pm1.8~\Lsun$, consistent with previous values \cite[e.g. $\lapp=25.6\;\Lsun$,][]{ monnier+10}.

The next quantity was the projected equatorial velocity $v\sin i$. We derived it from a series of 112 high resolution spectra collected on 2024, June 4th \cite[e.g.][]{rieutord+25b} with the Neo-Narval spectrograph \citep{boehm+16} attached to the 2-meter Bernard Lyot telescope at Pic du Midi. A mean line profile was obtained with the Least Square Deconvolution method \citep[LSD,][]{donati+97} applied to the observed spectra. The typical Signal-to-Noise ratio is $\sim$15,000 per resolved element. Fitting a synthetic, rotationally broadened, line profile, we could derive a precise value, namely $v\sin i=224.3\pm 2.6$~km/s. We show in Fig.~\ref{Rasal_Vsini} the median line profile and its fit. Uncertainty was estimated with different statistical methods such as bootstrapping and Monte-Carlo method.
We did not take into account the possible effects of micro- or macro-turbulence on the line profile. To our knowledge, there is no measurement of these parameters for Rasalhague. However, noting the similarity of this star with Vega (see conclusions), and using Vega's macro-turbulence estimate by \cite{hill+10}, we find a possible contribution of macro-turbulence on line profile of $\sim0.1$~km/s, which is negligible compared to present uncertainties.
Finally, let us point out that this line profile clearly shows the presence of the secondary\footnote{The best description of this star is given in \cite{gardner+21} who gave a mass for the secondary equal to $0.824\pm0.023\Msun$. At the age of Rasalhague ($\sim$600~Myrs), such a mass implies an effective temperature of 4750~K and a luminosity of 0.25$\Lsun$ according to CESAM2k20 models \citep{manchon+25}.} of the Rasalhague system as the little dip near the bottom of the line and redshifted by its relative radial velocity of $\sim+14$~km/s. At the time of the observations, the orbital phase of the binary was near 0.4, implying a radial velocity of $\sim+15.6$~km/s according to \cite{gardner+21}'s Fig.~3, which squares with our measure.

\begin{figure*}[t]
\sidecaption
\includegraphics[width=12cm]{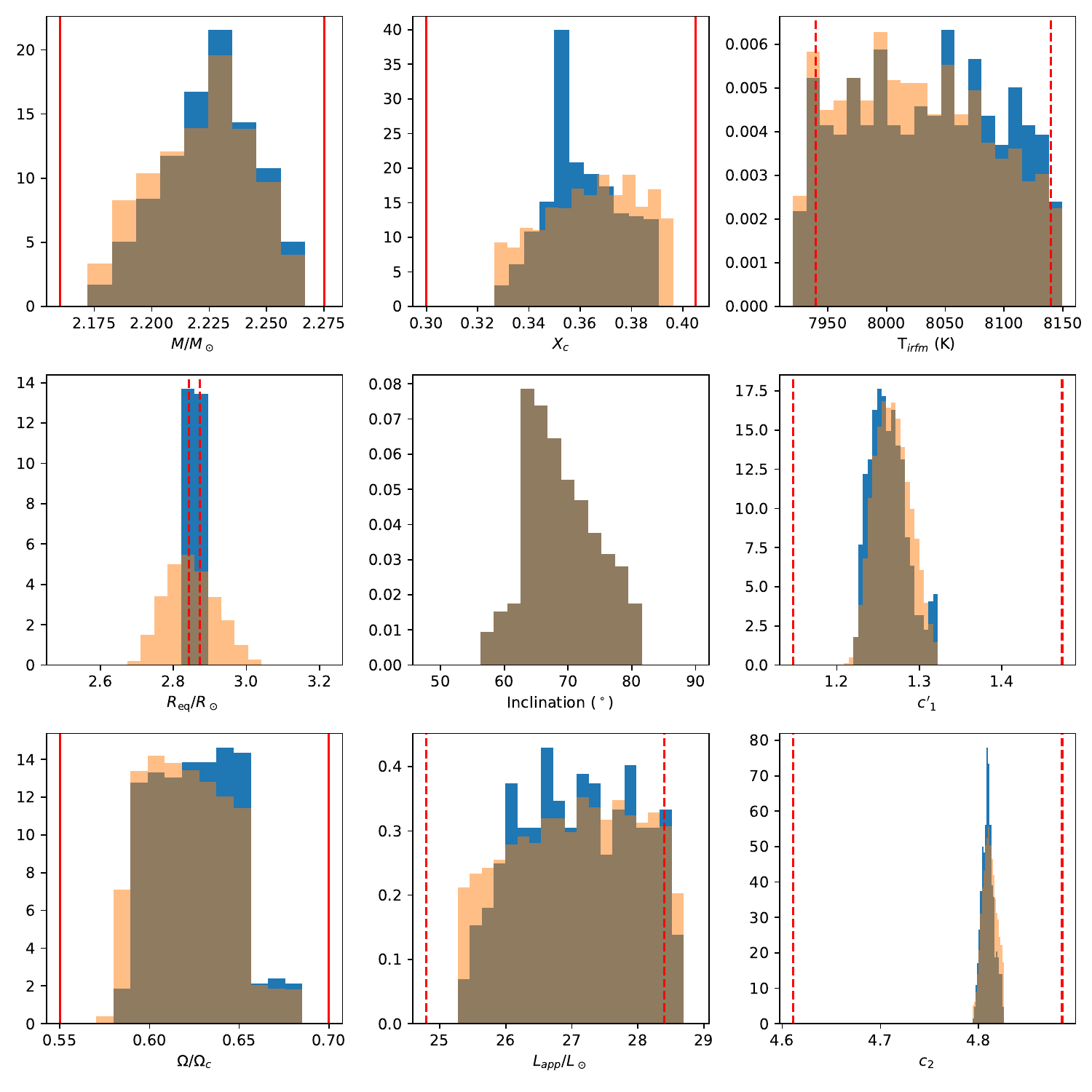}
\caption{Histograms of the distributions of mass $M$, rotation rate with respect to the critical one $\Omega/\Omega_c(=\omega)$, equatorial radius $R_{eq}$, relative hydrogen mass fraction in the core $X_c$, inclination of the rotation axis $i$, apparent luminosity $\lapp$ and spectroscopic quantities $\tIRFM$, $c'_1$ and $c_2$. The blue histograms show the distribution when the interferometric constraint on $R_{eq}$ from \cite{monnier+10} is used, while the superimposed orange ones show the distributions when no interferometric constraint is used. Red lines on mass, $\Omega/\Omega_c$ and $X_c$ show the interval of parameter that has been scanned. Dashed red lines on the observed parameters show the constraints that had to be matched.}
\label{figRasalHisto}
\end{figure*}
 
\begin{figure*}[t]
\centerline{
\includegraphics[width=\textwidth]{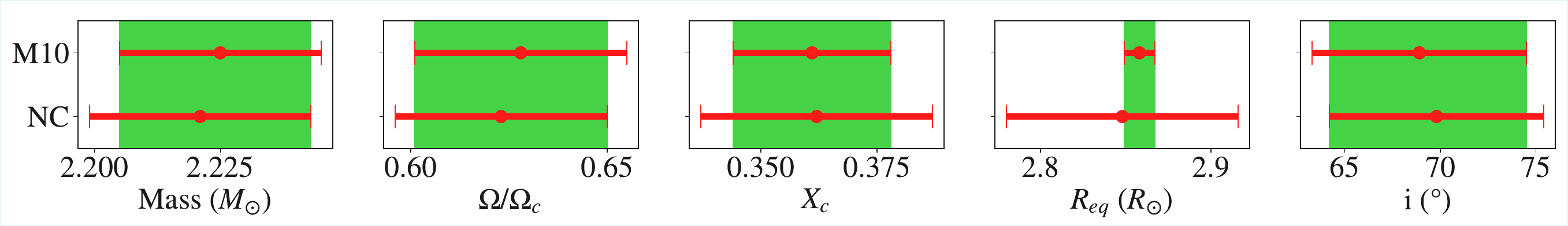}
}
\caption{Diagrams summarising the histograms of Fig.~\ref{figRasalHisto} when no constraint on the equatorial radius is applied (NC line) or when models meeting the equatorial radius of~\cite{monnier+10} are selected (M10 line). The red dots show the mean value and the segments the standard deviation. Green columns highlight the common values.}
\label{sumup}
\end{figure*}

The remaining constraints $\tIRFM$, $c'_1$ and $c_2$ were derived from spectrophotometry, with spectra of \cite{burnashev85} obtained through the VizieR service \citep{vizier}. Fluxes were provided for wavelengths between 320 and about 760~nm. Among the various spectra, we chose \#1401, whose Str\"omgren $c_1$ index (equal to 1.015) was closest to the value of \cite{paunzen15}, namely 1.039. This index was computed from each spectrum with a Python-translated adaptation of the \cite{Kurucz1993} codes calibrated with the CALSPEC\footnote{https://www.stsci.edu/hst/instrumentation/reference-data-for-calibration-and-tools/astronomical-catalogs/calspec} \citep[e.g.][]{bohlin+14,bohlin+22} spectrum of Vega. To compute the three observables defined previously, we needed to know the flux farther than 320~nm in the UV and farther than 760~nm in the near-IR domain. We completed the UV part of the spectrum with International Ultraviolet Explorer \citep[IUE,][]{IUE} data obtained through the MAST archive\footnote{https://archive.stsci.edu/iue/}. From the so-called long-wavelength IUE data, we selected the three large aperture spectra that contained no negative flux values (LWR12747LL, LWR14250LL, and LWR14482LL). We then considered an average flux obtained with a mean weighted by the square root of the exposure time. For the IR side, needed for computing $\tIRFM$, we used a synthetic spectrum based on the ESTER model that matched best the data of \cite{monnier+10}, setting the inclination to $i=90^\circ$.

Each extension has been scaled so as to ensure continuity with the \cite{burnashev85} spectra, the mean IUE spectrum being degraded to the resolution of the spectrophotometric data beforehand. The scaling factor of the IUE part was of the same order of magnitude as the calibration correction proposed by \cite{bohlin_bianchi18}. We ended up with the spectrum plotted in Fig.~\ref{figRasalSpectres}, from which we computed the observables displayed in Table~\ref{Rasal_obs}\footnote{Note that in Fig. 15 of \cite{lazzarotto+23} the plotted bandpass profile of the F220 filter was erroneously that of the ACS\_HRC.F220W instead of that of the HSP\_UV1\_F220W\_A, which we used in both works.}. We used a conservative value of 5\% for the flux uncertainties \citep{glushneva+92}, which translated into 0.11 mag for the $c'_1$ and $c_2$ indices, of the same order of magnitude as that of typical $c_1$ photometric measurements \citep{Crawford_Barnes1970}. This systematic error was combined with the dispersion between the IUE spectra to yield the global uncertainty reported in Tab.~\ref{Rasal_obs}.

\begin{table}[t]
\caption{{\it Top:} Most probable parameters of Rasalhague with 2D-ESTER models using the interferometric constraint on equatorial radius by  \cite{monnier+10}. {\it Bottom:} the 2D-ESTER concordance model of Rasalhague that fits all the observed quantities. The M10 column shows values derived by \cite{monnier+10}.}
\label{Final_Rasalhague}
\begin{center}
\begin{tabular}{l r}
\hline
\hline
Stellar parameter  &  Our derivation\\
\hline
$i\; (^\circ)$  & $68.9\pm5.6$ \\
$M\; (M_\odot)$ & $2.225 \pm 0.021 $\\
$\lapp\; (L_\odot)$ & $27.08 \pm 0.87 $\\
$R_\mathrm{eq}\; (R_\odot)$ & $2.858 \pm 0.009$ \\
$\Omega_\mathrm{eq}/\Omega_c$ & $0.628 \pm 0.026$ \\
$X_\mathrm{core}/X_\mathrm{env}$& $0.361 \pm 0.016$ \\
$T_\mathrm{IRFM}\; (\mathrm{K})$&$8035\pm63$ \\ 
$c_1'$&$1.310\pm0.163$\\ 
$c_2$& $4.748\pm0.137$\\
\hline
\end{tabular}
\begin{tabular}{lcc}
\multicolumn{3}{c}{Best Rasalhague model} \\
Parameters & This work & M10 \\
$M\; (M_\odot)$ & 2.225 & $2.18\pm0.02$\\
$L\; (L_\odot)$ & 29.3 & $31.3\pm1$\\
$R_\mathrm{pole}\; (R_\odot)$ & 2.383& $2.388 \pm 0.013 $\\
$R_\mathrm{eq}\; (R_\odot)$ & 2.859& $2.858 \pm 0.015$ \\
$\epsilon=1-R_\mathrm{pole}/R_\mathrm{eq}$ & 0.167& 0.164 \\
$\teff^\mathrm{pole}\;$ (K) & 9051 & $9384 \pm 154$ \\
$\teff^\mathrm{eq}\; $(K) & 7609 & $7569 \pm 124$\\
V$_\mathrm{eq}$ \; (km/s) & 242 \\
$P_\mathrm{rot}^\mathrm{pole}$ (hr) & 14.76&  \\
$P_\mathrm{rot}^\mathrm{eq}$(hr) & 14.33&$ 14.55\pm0.35$ \\
$X_\mathrm{env}$ & 0.72 \\
$X_\mathrm{core}$ & 0.260\\
$Z$ & 0.019\\
$i (^\circ)$ & 69\\
$\lapp$ ($L_\odot$) & 26.9 & 25.6\\
\hline
\hline
\end{tabular}
\end{center}
\end{table}
 
\subsection{Fundamental parameters of Rasalhague}

As explained in Sect.~\ref{models}, we selected among a grid of 8162
models those matching the foregoing observational constraints. The
parameters of the grid models were set around the values of the initial ESTER
model that matched best the previous estimates by \cite{monnier+10},
as reported in Table~\ref{Rasal_param}. We thus scanned a mass range
from 2.16 to 2.275~$\Msun$, a rotation rate from 0.55 to 0.7, and a
relative hydrogen mass fraction in the core, $X_c$ from 0.3 to 0.4. The
mass fraction of hydrogen in the envelope $X_{\rm env}$ was fixed to the
initial solar one X=0.72 \cite[e.g.][]{AGSP09}, which seems reasonable
for a star much younger than the sun. The metallicity was computed from
the relative abundances of \cite{erspamer_north03}, using solar abundances
of \cite{GS98}, thus yielding Z=0.019.

We show in Fig.~\ref{figRasalHisto} the histograms of the parameters when
either only the five observables $\lapp, v\sin i, \tIRFM, c'_1$ and $c_2$
were used, or when the interferometric measure of $R_\mathrm{eq}$ from
\cite{monnier+10} was added. Quite remarkably, the spectrophotometric
equatorial radius agreed well with the interferometric value. The
error bar on the interferometric value is however much less, as
expected. However, the added value of the spectrophotometric method
is the determination of the rotation axis inclination on the line of
sight. \cite{zhao+09} could not determine this parameter properly as it
turned out to be degenerate with the gravity darkening exponent\footnote{A
simplified model of gravity darkening assumes that the latitudinal
variations of effective temperature follow the latitudinal variations of
the effective gravity as $\teff\propto g_\mathrm{eff}^\beta$. However,
\cite{ELR11} have shown that $\beta$ depends on the rotation rate of
the star.} $\beta$. Because of this degeneracy, \cite{zhao+09} reverted
to $\beta=0.25$, the value derived by \cite{vonzeipel24a}, which is
valid only for small rotation rates \citep{ELR11}. This assumption led
\cite{zhao+09} to impose an inclination of 88.5$^\circ$, slightly revised
to $87.5^\circ$ by \cite{monnier+10}, but still much larger than ours.

Besides, Rasalhague belongs to a binary system, and \cite{gardner+21}
measured an orbital plane inclination of $130.679\pm0.067^\circ$ with
respect to the line of sight. Because spectroscopy and interferometry are
blind on the direction of Rasalhague's rotation, our results indicate that
the spin vector can be oriented either at 70 or 110$^\circ$ with respect
the line of sight. Consequently, the mutual orientation of the orbital and
spin angular momenta are either 20 or 60$^\circ$ apart. Hence, according
to present data, it seems difficult to assess that the two angular
momentum vectors are aligned, but histograms of Fig.~\ref{figRasalHisto}
leave the possibility that they be just 5 to 10$^\circ$ apart.

To take the best of spectrophotometric and interferometric measurements we
also ran our model selection with spectrophotometric constraints together
with the bounds on the equatorial radius given by \cite{monnier+10}. As
shown in Fig.~\ref{figRasalHisto}, the distributions were marginally
altered, except that of $R_\mathrm{eq}$ of course. They also showed
that the two photometric indices, $c'_1$ and $c_2$, do not participate
in the selection of the models in the present case. As a side note,
we also remark that histograms show that some models are outside the
uncertainty interval of observed parameters. This is typical of models
for which a change of inclination, within its uncertainties, put them
in the observed range.

To be complete, we also used the radius determination of \cite{baines+18}, which was derived with the limb-darkened disc model. 
The resulting new distributions (not shown) presented only slight differences with those displayed in Fig.~\ref{figRasalHisto}. In particular, the average inclination remained close to 70$^\circ$.

We summarised our results on Rasalhague in Fig.~\ref{sumup}, which
shows that the models converge to a rather well defined set of
fundamental parameters.  In Table~\ref{Final_Rasalhague}, we have
gathered all the values derived from our analysis with their error
bars. Our mass agrees pretty well with the value of \cite{gardner+21},
$2.20\pm0.06 M_\odot$, which is a dynamical mass derived from the orbit
of Rasalhague's companion. We used the mass, the rotation rate $\omega$
and the ratio $X_c=X_\mathrm{core}/X_\mathrm{env}$ to design a new ESTER
model, which is our concordance model for Rasalhague. In this same
table we also report the values found by \cite{monnier+10}, who used
a Roche model for their determination. Their results are quite close
to ours, except for the effective temperatures, likely because of the
discrepancy on the inclination. In this table we also report the
polar and equatorial rotation period. This is an output of the model,
which includes the baroclinicity of the radiative envelope leading to
the differential rotation of the star. It gives an idea of the expected
surface differential rotation of Rasalhague.

As a final touch, we computed a synthetic spectrum of Rasalhague based
on the parameters of Tab.~\ref{Final_Rasalhague}. This spectrum is
plotted along the observed one in Fig.~\ref{figRasalSpectres}. Lines
have been thickened with a greyed or greened over line in order to show
the uncertainties of both the observed and synthetic spectra. For the
synthetic spectrum we just considered the effect of inclination error
bar. We see that the observed SED and the synthetic one match fairly well.

\begin{figure}[t]
\includegraphics[width=1.0\linewidth]{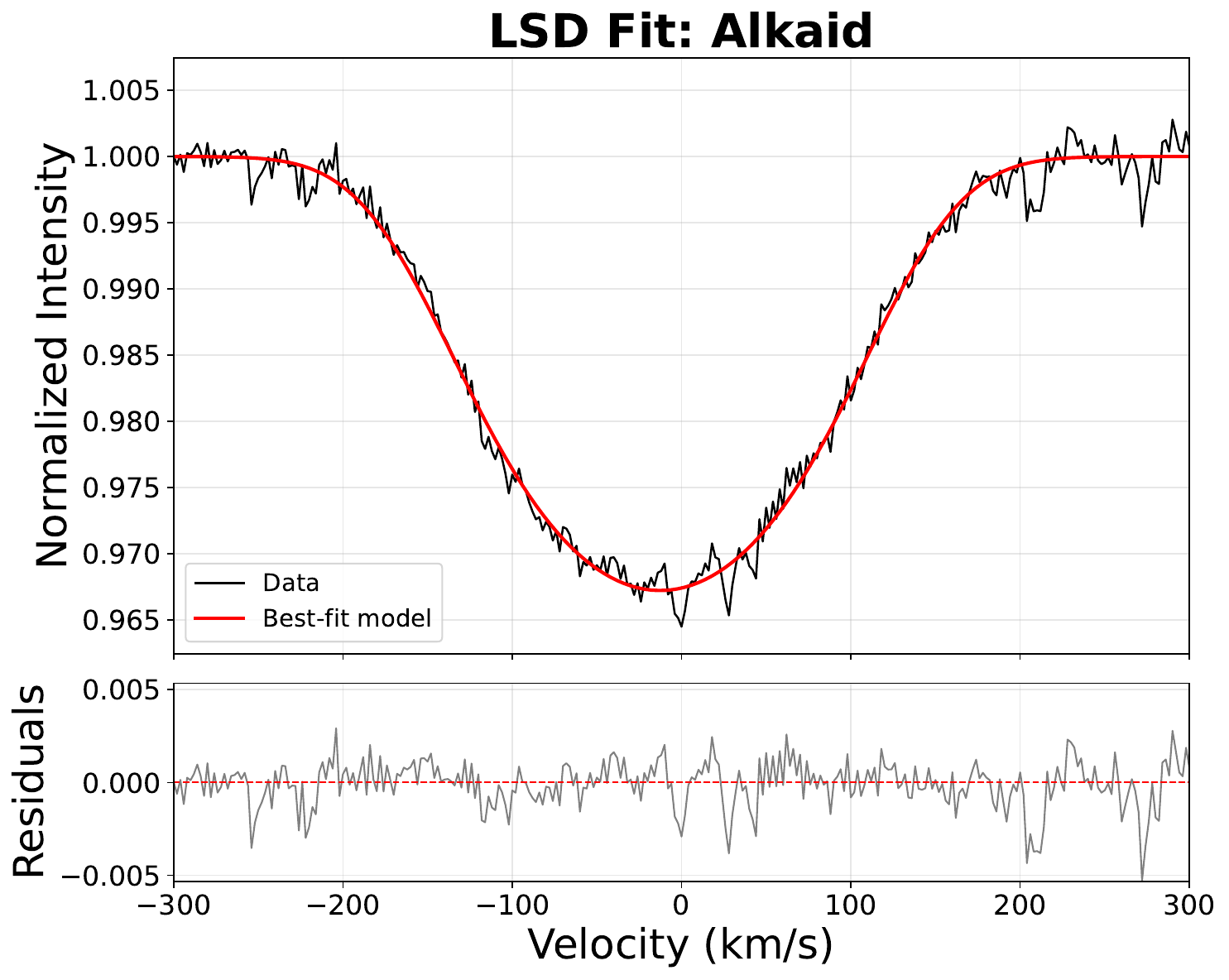}
\caption{Mean line profile of Alkaid obtained with the Least Square
Deconvolution method with a rotationally broadened profile fit to the data (top panel), and its residuals (bottom panel). The resulting projected velocity is $154.3\pm9.1$~km/s.}
\label{Alkaid_vsini}
\end{figure}

To be complete, we also wished to provide an estimate of the age
of Rasalhague. To this end, we ran 1D evolving models until $X_c$
reached a value of 0.361. We run the CESAM2k20 \citep{manchon+25},
the Toulouse-Geneva evolution codes \citep[][]{hbh08}
and the ``Modules for Experiments in Stellar Astrophysics"
\citep{paxton+11,paxton+13,paxton+15,paxton+18,jermyn+23}. They all
give an age around 600~Myrs past the ZAMS, in agreement with the
previous estimate of \cite{monnier+10} who used another 1D evolution
code, namely $Y^2$ of \cite{demarque+04}. However, the foregoing age
was obtained with models neglecting any ``turbulent" diffusion. When
such a diffusion is included the convective core is fueled with fresh
hydrogen and it takes more time to reach a given $X_c$. Some tests
showed us that the increase can be significant, namely a few tens of
percent. We thus conclude that 600~Myrs is certainly a lower limit to the
real age of Rasalhague. Hence, the precise estimate of the age of this
star will have to wait for 2D models including turbulent transport to be
operational\footnote{2D-ESTER evolution code faces numerical difficulties
in the mass range of Rasalhague.}.

\begin{figure*}[t]
\sidecaption
\includegraphics[width=12cm]{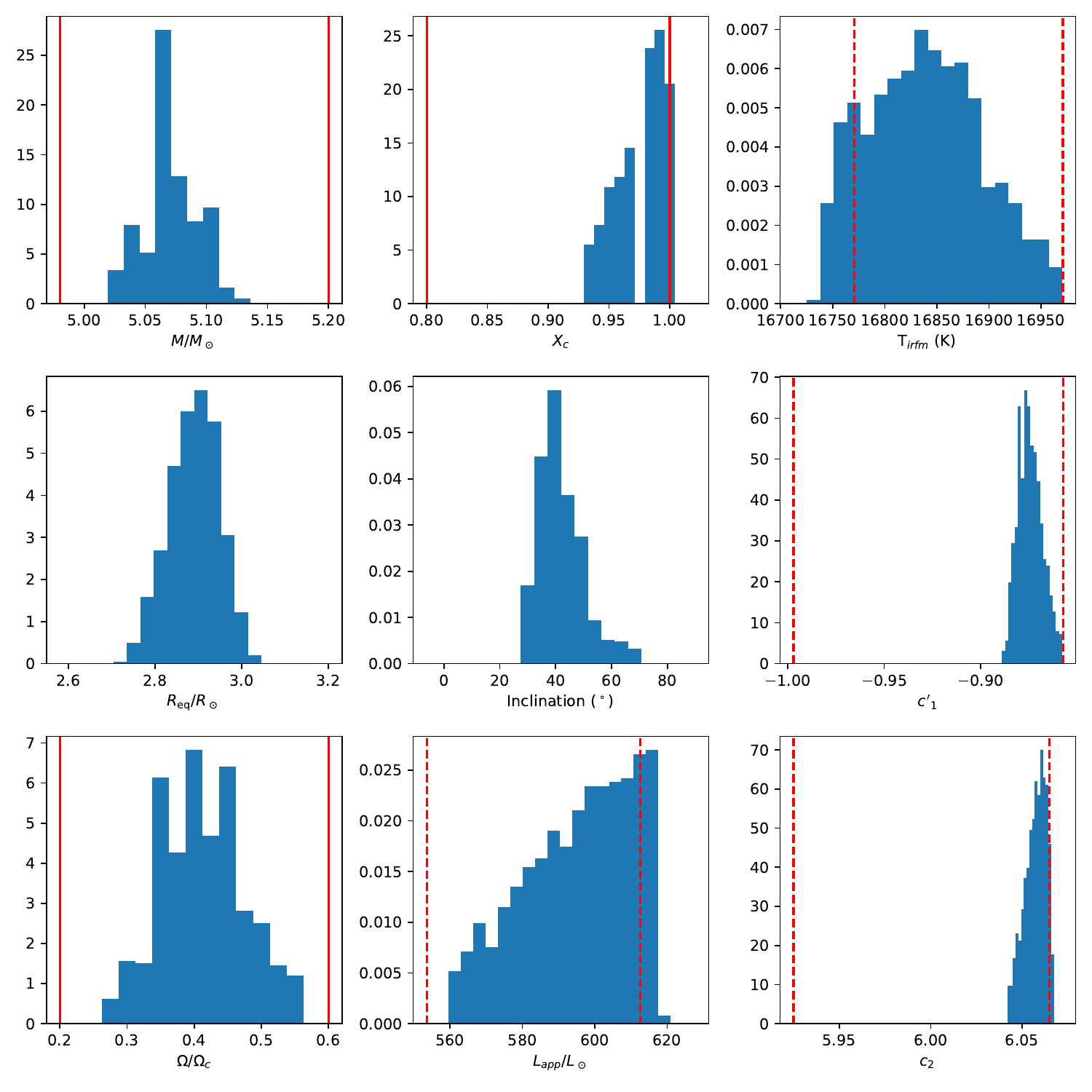}
\caption{Histograms of the parameter distributions of steady 2D-ESTER models for Alkaid that match the observational constraints. Red lines have the same meaning as in Fig.~\ref{figRasalHisto}.}
\label{figHistosAlkaid}
\end{figure*}

\section{Alkaid -- $\eta$ UMa}\label{alkaid}
\subsection{Observational constraints}

We used the CALSPEC spectrum\footnote{The file is {\tt
etauma\_stis\_008.fits}.}  of Alkaid to derive the observables
we needed: $c'_1=-0.927\pm0.07$, $c_2=5.995\pm0.07$, and
$T_\mathrm{IRFM}=16781\pm100$~K. The uncertainties were determined
from the flux accuracy in each part of the spectrum. We determined
the bolometric flux of Alkaid on Earth by integration of the CALSPEC
spectrum, and using the Hipparcos parallax, from which we found a distance
of $31.38 \pm0.24$~pc. The subsequent apparent luminosity was equal
to $L_\mathrm{app}=583\pm30~L_\odot$, in agreement with the value of
\cite{baines+18}, who quoted $594\pm31~L_\odot$. We did not consider the
revised values of \cite{baines+25} because of the much lower effective
temperature ($13946 \pm 60$~K) they obtained, inconsistent with the
values we derived from CALSPEC data.

The value of $v \sin i$ has been computed the same way as for Rasalhague but with spectra extracted from PolarBase\footnote{https://www.polarbase.ovgso.fr/}, a database for high resolution spectropolarimetric data \citep{petit+14}. We extracted
52 spectra collected on 2015, March 12th, with the Narval
spectropolarimeter \citep{auriere03}. The Signal-to-Noise ratio of the resulting LSD profile is $\sim2500$ per resolved element.
The best fit yielded $v \sin i=154.3\pm9.1$~km.s$^{-1}$. We show in Fig.~\ref{Alkaid_vsini} the actual line profile and its adjustment by a rotationally broadened absorption line. These observational constraints are gathered in Table~\ref{Alkaid_obs}.

\begin{table}[t]
\caption{Observational constraints for Alkaid.}
\label{Alkaid_obs}
\begin{center}
\begin{tabular}{lc}
\hline
\hline
$\tIRFM\ (\mathrm{K})$&16871$\pm100$\\
$c_1'$&-0.927$\pm0.07$\\
$c_2$&5.995$\pm0.07$\\
$L_\mathrm{app}\ (L_\odot$)&$583\pm30$\\
$v\sin i$\ (km/s)&   $154.3\pm9.1$\\
\hline
\hline
\end{tabular}
\end{center}
\end{table}

\subsection{Parameters derived from the models}

\begin{table}[t]
\caption{Top: Most probable parameters of Alkaid with steady 2D-ESTER models. Bottom: the 2D-ESTER concordance model of Alkaid that fits all the observed quantities.}
\label{Alkaid_results}
\begin{center}
\begin{tabular}{l r}
\hline
\hline
Stellar parameter  &  Our derivation\\
\hline
$i\; (^\circ)$  & 41.9$\pm$8.2 \\
$M\; (M_\odot)$ & 5.071 $\pm$ 0.023 \\
$\lapp\; (L_\odot)$ & 593.7 $\pm$ 16.8 \\
$R_\mathrm{eq}\; (R_\odot)$ & 2.891 $\pm$ 0.057 \\
$\Omega_\mathrm{eq}/\Omega_c$ & 0.413 $\pm$ 0.062 \\
$X_\mathrm{core}/X_\mathrm{env}$& 0.971 $\pm$ 0.024 \\
$T_\mathrm{IRFM}\; (\mathrm{K})$&$16841\pm37$ \\
$c_1'$&$-0.874\pm0.006$\\
$c_2$& $ 6.057\pm0.006$\\
\hline
\end{tabular}
\begin{tabular}{lcc}
\multicolumn{2}{c}{Best Alkaid model} \\
Parameters & This work \\
$M\; (M_\odot)$ & 5.085 \\
$L\; (L_\odot)$ & 574  \\
$R_\mathrm{pole}\; (R_\odot)$ & 2.665\\
$R_\mathrm{eq}\; (R_\odot)$ & 2.894\\
$\epsilon=1-R_\mathrm{pole}/R_\mathrm{eq}$ & 0.079\\
$\teff^\mathrm{pole}\;$ (K) &17720\\
$\teff^\mathrm{eq}\; $(K) &16329 \\
V$_\mathrm{eq}$ \; (km/s) & 240 \\
$P_\mathrm{rot}^\mathrm{pole}\;$ (hr) & 15.6   \\
$P_\mathrm{rot}^\mathrm{eq}\; $(hr) & 14.6  \\
$X_\mathrm{env}$ & 0.70 \\
$X_\mathrm{core}$ & 0.679\\
$Z$ & 0.020\\
$i\; (^\circ)$ & 42\\
$\lapp$ ($L_\odot$) & 596  \\
\hline
\hline
\end{tabular}
\end{center}
\end{table}

To start our grid of models we first observed that the apparent luminosity of Alkaid could be reproduced by a ZAMS non-rotating 5.1$\Msun$ ESTER model (L=583~$\Lsun$) with X$_{\rm env}=0.7$ and Z=0.02. We kept this chemical composition throughout the analysis of Alkaid. Its radius of 2.65~$\Rsun$ combined with a minimum equatorial velocity equal to the observed $V\sin i$ gave a minimum value $\omega\sim0.28$. Assuming that the star is young, we therefore scanned the following parameter space:
\beqa
&& 4.98\leq M/\Msun \leq 5.20 \\
&& 0.2\leq \omega \leq 0.6 \\
&& 0.8\leq X_c \leq 1
\eeqa
where we distributed about 5100 models.

We then selected models matching the observational constraints as for Rasalhague, but here no measure of equatorial radius was used. Some interferometric measurements of the angular diameter have been done by \cite{baines+18,baines+25,gordon+19} and \cite{abe+24}, but these works turned out to be too imprecise to be useful on the selection of models.

\begin{figure}[t]
\includegraphics[width=1.0\linewidth]{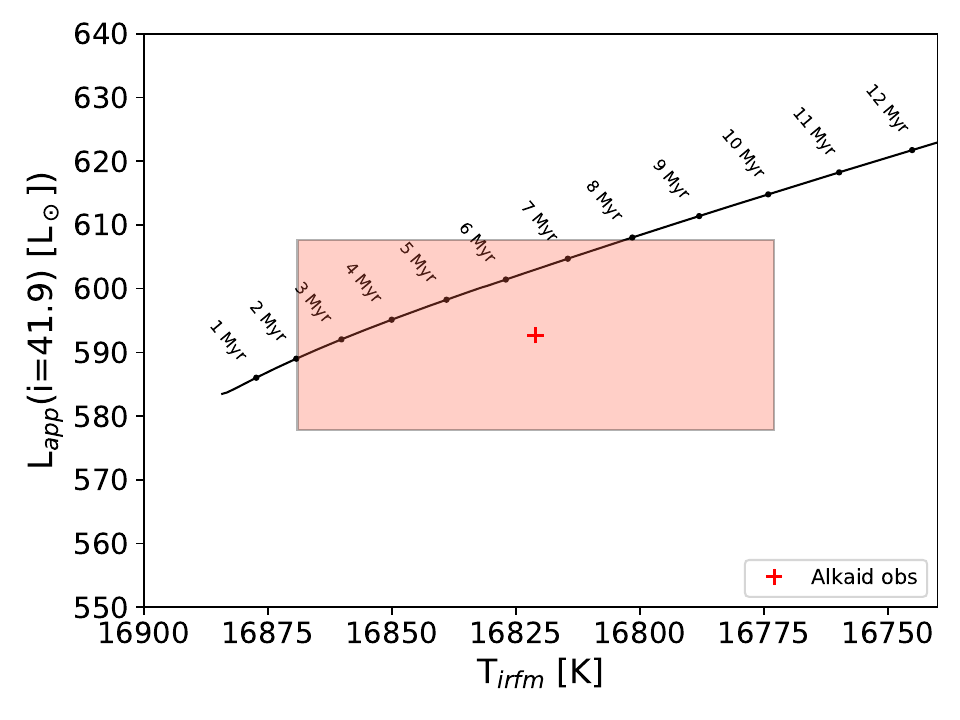}
\caption{Evolutionary track in a Hertzsprung-Russell-like diagram showing the evolution the best ESTER Alkaid model during the first 12 Myrs on the main sequence. The red cross represents the mean observed value and its uncertainty box is depicted by the pink area.}
\label{Evol_Alkaid}
\end{figure}

The parameters and observables histograms for Alkaid are displayed in Fig.~\ref{figHistosAlkaid}. As shown by the red lines on the scanned parameters ($M,\Omega/\Omega_c$, and $X_c$), the grid was wide enough not to miss any possible model. Clearly, the four observables $\lapp$, $\tIRFM$, $c'_1$ and $c_2$ played a role in the selection of models, and the most probable parameters were clearly emerging.
We gathered in Table~\ref{Alkaid_results} all the parameters we determined, along with those of the concordance ESTER model. The proxy of the age $X_c$ points to 0.97, meaning that Alkaid is barely off the ZAMS. To put a figure for the age, we run a time-evolving 2D model with the ESTER code. We plotted the evolutionary track of the star in the ($\tIRFM,\lapp$) plane in Fig.~\ref{Evol_Alkaid}, which shows that Alkaid has evolved for $6^{+2}_{-4}$~Myrs after the ZAMS.

From the foregoing results and the distance to this star we were also
able to infer its equatorial angular diameter, which we find equal to
0.844$\pm0.009$~mas (uncertainty comes from that of the distance). This
value is consistent with $0.981\pm0.144$~mas of \cite{baines+18},
with $\theta_{LD}=0.834\pm0.06$~mas of \cite{gordon+19} and with
$\theta_{LD}=0.828\pm0.029$~mas of \cite{abe+24}. These latter
measurements have been obtained with interferometry (intensity
interferometer for \cite{abe+24}) and do not distinguish between polar and
equatorial radii. If we assume that they measure the radius of a disc of
surface equal to the projected surface of the star on the sky, our model
says that the angular diameter of that disc would be 0.826$\pm0.006$~mas,
still in line with the foregoing observations.

\section{Conclusions}\label{conclusion}

We showed that the fundamental parameters of early-type fast rotating stars can be derived from high quality spectrophotometric data, and high-resolution high-signal-to-noise ratio spectroscopy. In addition, we were able to determine the inclination of their rotation axis on the line of sight. This is possible thanks to gravity darkening, which makes the characteristics of the spectral energy distribution dependent on the orientation of the rotation axis and on the rotational flattening. This allowed us to infer the parameters for two stars: Rasalhague ($\alpha$ Oph) and Alkaid ($\eta$ UMa). For the first star we could derive the true orientation of its rotation axis, a quantity that escaped the interferometric measurements because of a degeneracy between this angle and the so-called gravity darkening exponent \citep{zhao+09,monnier+10}. The choice of Alkaid was directed by the availability of a complete spectral energy distribution from CALSPEC, hence well calibrated data from space instruments covering the spectral interval from the ultraviolet to the near infrared domains. This star, of spectral type B3V, also offered a test of our method for the case of hot stars.

For now, our method uses spectrophotometric data because fluxes in the UV bands we defined are not available. Should some UV photometry with bands near those we used be available someday, the method would only require the different magnitudes in the filters used to compute the $c_1'$ and $c_2$ indices. As for $\tIRFM$, we would compute it from an IR flux and the bolometric flux. The case of stars far enough for reddening to be considered could thus be treated, provided we have dereddening laws for the $c_1'$ and $c_2$ indices, as they exist for the $uvby\beta$ system.

Of course, one has to keep in mind that our method has its limits: $(i)$ it only works if the gravity darkening is strong enough, hence if rotation is fast enough. The actual limit is not known, but we estimate it at 30\% of the critical angular velocity, value that has to be checked. $(ii)$ we need data in the UV bands, hence from space. $(iii)$ the parameters we derived represent an element of a set of ESTER models. Obviously, if we change the physics or some assumed inputs of the models, the parameters we derive are likely to change. For instance the chemical composition was assumed and fixed. For our method to provide realistic results, we need at least some clues about the actual composition.

The positive side of these results is that we now know better these two stars. About Rasalhague, we basically confirm the results \cite{monnier+10} obtained with interferometry and seismology, and we complete them by giving the inclination of its rotation axis on the line of sight ($69\pm6^\circ$). Such an inclination, together with rotation, could also be constrained by linear polarization measurements like Regulus ($\alpha$ Leo) as worked out by \cite{cotton+17}.

We note that Rasalhague is quite similar to Vega in mass and age, but its spectral type A5IV, instead of A0V for Vega, essentially reflects its nearly equatorial view compared to the nearly polar-on view of Vega. This result shows that the determination of the spin axis orientation is essential for rapidly rotating stars, when one interprets the spectral type in terms of evolutionary status. With Rasalhague and Vega  we have an example where the former is younger than the latter, although of later spectral type (and higher luminosity class). 

The case of Alkaid showed that our method also works on such hot
intermediate-mass stars. We learned that it is a very young star,
typically 6~Myrs off the ZAMS, with a mass $\sim5.07\ \Msun$.

An interesting consequence of knowing or better knowing the inclination
of the rotation axis on the line of sight is that we can now determine
more accurately the visibility of the oscillation modes of these two
stars. Rasalhague is well known as a $\delta$-Scuti oscillator, with a
rich oscillation spectrum \citep{monnier+10}, while Alkaid also shows
oscillations \citep{rudrasingam+26}, the spectrum of which displays a
peak at period about 14.1~hrs, which could be associated with the rotation
period of this star, since we find it at P$_{rot}= 14.6\pm1.9$~hrs.

A natural follow up of this work will be refinements of the proposed
models of these two stars, taking into account the seismological
constraints from the TESS satellite like it has been done with Altair
\citep{bouchaud+20,rieutord+24}. This will be presented in a future work.

\begin{acknowledgements}
We are very grateful to the referee for many valuable comments, which
helped us improve the original manuscript.
The research leading to these results has received funding from the
European Research Council (ERC) under the Horizon Europe programme
(Synergy Grant agreement N$^\circ$101071505: 4D-STAR).  While partially
funded by the European Union, views and opinions expressed are however
those of the authors only and do not necessarily reflect those of the
European Union or the European Research Council.  Neither the European
Union nor the granting authority can be held responsible for them. Most
of the calculations have been performed on CALMIP supercomputing center
(Grant 2023-P20025).
This research made also use of the VizieR catalogue access tool, CDS,
 Strasbourg, France (DOI : 10.26093/cds/vizier).
\end{acknowledgements}

\bibliographystyle{aa}
\bibliography{bibnew}

\end{document}